\documentclass[12pt]{elsarticle}
\usepackage{amsmath}
\usepackage{amssymb}
\usepackage[cp1251]{inputenc}
\usepackage[russian]{babel}

\usepackage{latexsym}

\begin{document}

\begin{frontmatter}

\title{
 Group analysis of the Fourier transform of the spatially homogeneous and isotropic Boltzmann equation
 \\
  with a source term.
}

\author{A. Suriyawichitseranee}

\address{School of Mathematics, Institute of Science, Suranaree University
of Technology,
\\
 Nakhon Ratchasima, 30000, Thailand}

\ead{amornrat@math.sut.ac.th}

\author{Yu.N. Grigoriev}

\address{Institute of Computational Technology, Novosibirsk, 630090, Russia }

\ead{grigor@ict.nsc.ru}

\author{S.V. Meleshko}

\address{School of Mathematics, Institute of Science, Suranaree University
of Technology,
\\
Nakhon Ratchasima, 30000, Thailand}

\ead{sergey@math.sut.ac.th}

\begin{abstract}
The  paper is devoted to group analysis of the spatially homogeneous
and isotropic Boltzmann equation with a source term. In fact, the
Fourier transform of
the Boltzmann equation with respect to the molecular velocity variable  is considered. Using a particular class of
solutions, the determining equation for the admitted Lie group is reduced to a partial
differential equation for the source function. The latter equation
is analyzed by an algebraic method. A complete group
classification of the Fourier transform of the Boltzmann equation
is given. All invariant solutions of this equation are
also presented in the paper.
\end{abstract}

\end{frontmatter}

\section{Introduction}

The classical Boltzmann equation plays a central role in kinetic gas theory.
At the same time, there is frequently the need to include additional source terms into it.
In particular, this is applied in the so-called  "extended kinetic theory"
\cite{bk:BoffiSpiga[1982a],bk:BoffiSpiga[1982c]}, where source terms which are linear with respect to the distribution function are appended to the Boltzmann equation.
The Boltzmann equation with autonomous sources, independent of the distribution function,
are of interest for the kinetic description of initiation of high threshold processes by
"hot" particles, for reacting  gas flows over catalytic surfaces, and others. A first attempt of application of group analysis for studying the Boltzmann equation with such a source term was
done in \cite{bk:Nonnenmacher[1984]}, where by using the method of \cite{bk:KW}, the homogeneous and
isotropic Boltzmann equation with the Maxwell molecular model was reduced to the equation for
a generating function of power moments. Complete Lie
symmetries of the equation for the moment generating function were recently found in
\cite{bk:GrigorievMeleshkoSuriyawichitseranee[2014]}. However, the 
transformation of the invariant solutions thus obtained into the in corresponding
solutions of the original Boltzmann equation is a very complicated task. An example
of typical difficulties can be found in \cite{bk:KW}, where this transition obstacle  was
overcome for obtaining the BKW-solution.

A more promising approach consists of considering the Fourier image
equation of the homogeneous and isotropic Boltzmann equation
\cite{bk:Bobylev[1975a]}, on the base of which the BKW-solution  was directly derived in
\cite{bk:Bobylev[1976]}. 
The complete group analysis of this
equation without a source term was completed in \cite{bk:GrigMel[1987]}.
In the present paper these results are generalized
for the non-uniform case.

The paper is organized as follows.
After introducing the equation studied, the equation defining an admitted Lie group is formulated.
Solution of the determining equation is reduced to the study of a partial differential equation with some undefined constants. For convenience of a reader this part of the study is presented in Appendix. For further study of the reduced equation an equivalence Lie group is obtained. It is shown that actions of the obtained equivalence transformations are similar to the automorphisms of the considered four generators. This allows us to use a method of constructing an optimal system of subalgebras for the group classification. These studies are followed by consideration of all possible invariant solutions of the Fourier image equation of the homogeneous and isotropic Boltzmann equation with a source term.

\section{The equation studied}

The Fourier image of the spatially homogeneous and isotropic Boltzmann equation
with a source term has the form \cite{bk:Bobylev[1975a]}:
\begin{equation}\label{BEs}
\varphi _t(x,t)+\varphi (x,t)\varphi (0,t)=\int_0^1\ \varphi
(xs,t)\varphi (x(1-s),t)\ ds + \hat{q} (x,t).
\end{equation}
Here the function $\varphi(x,t)$ is related with the Fourier
transform $\tilde{\varphi}(k,t)$ of the  distribution function
$f(v,t)$, isotropic in the 3D-space of molecular velocities  by the
formulae
\[
\varphi(x,t)\equiv\varphi(k^2/2,t)=\tilde{\varphi}(k,t) =
{\frac{4\pi }{k}} \int_0 ^\infty v \sin (kv)f(v,t)\,dv.
\]
Similarly, the transform of the isotropic source function $q(v,t)$
is
\[
\tilde{\hat{q}} (k,t)= {\frac{4\pi }{k}} \int_0 ^\infty v \sin (kv)q(v,t)\,dv,
\]
and
\[
\hat{q} (x,t)\equiv \hat{q} (k^2/2,t) = \tilde{\hat{q}} (k,t).
\]

The inverse Fourier transform of $\tilde{\varphi} (k,t)$ gives the distribution function
\[
f(v,t) = {\frac{4\pi }{v}} \int_0 ^\infty k \sin (kv)\tilde{\varphi}(k,t)\,
dk.
\]

In the process of solving the determining equation, we use the property
that for any smooth function $\varphi_0(x)$ there exists a solution
of the Cauchy problem of equation (\ref{BEs}) with the initial data
\[
\varphi(x,t_0)=\varphi_0(x).
\]

\section{Determining equation}

The classical group analysis method cannot be applied to equation (\ref{BEs}).
One needs to use the method developed for equations with non-local
terms \cite{bk:GrigMel[1986],bk:Meleshko[2005],bk:GrigorievIbragimovKovalevMeleshko2010} instead.
A  generator of the admitted Lie group is sought in the form
\[
X=\xi(x,t,\varphi)\partial_{x}+\eta(x,t,\varphi)\partial_{t}
+\zeta(x,t,\varphi)\partial_{\varphi}.
\]
According to the algorithm, the determining equation for equation (\ref{BEs}) is
\begin{equation}\label{DEeq}
D_t \psi(x,t)+\psi(0,t) \varphi(x,t) + \psi(x,t) \varphi(0,t)-2\int_0^1 \varphi(x(1-s),t) \psi (xs,t) ds =0,
\end{equation}
where $D_t$ is the total derivative with respect to $t$, and the function $\psi(x,t)$ is
\[
\psi(x,t) = \zeta (x,t,\varphi(x,t)) - \xi (x,t,\varphi(x,t))\varphi_x(x,t) - \eta (x,t,\varphi(x,t))\varphi_t(x,t).
\]
The determining equation (\ref{DEeq}) has to be satisfied for any solution of equation (\ref{BEs}). This allows us to exclude the derivatives $\varphi _t$, $\varphi _{xt}$ and $\varphi _{tt}$ from the determining equation. In fact, differentiating (\ref{BEs}) and substituting $\varphi _t$ found from (\ref{BEs}), one obtains
\[
\begin{split}
\varphi _{xt}(x,t)&=-\varphi_x (x,t)\varphi (0,t)+2\int_0^1 s\varphi_x
(xs,t)\varphi (x(1-s),t)\ ds + \hat{q}_x (x,t), \\
\varphi _{tt}(x,t)&=\varphi (x,t)\varphi^2 (0,t)-3\varphi (0,t)\int_0^1 \varphi
(xs,t)\varphi (x(1-s),t)\ ds \\
&+2\int_0^1 \int_0^1 \varphi (x(1-s),t)\varphi
(xss',t)\varphi (xs(1-s'),t)ds'ds \\
&-\hat{q}(0,t)\varphi (x,t)-\hat{q}(x,t)\varphi (0,t)+2\int_0^1 \varphi (x(1-s),t)\hat{q}(xs,t)ds+ \hat{q}_t (x,t).
\end{split}
\]
\[
\begin{split}
\varphi _{xt}(x,t)&=-\varphi_x (x,t)\varphi (0,t)+2\int_0^1 s\varphi_x
(xs,t)\varphi (x(1-s),t)\ ds + \hat{q}_x (x,t), \\
\varphi _{tt}(x,t)&=\varphi (x,t)\varphi^2 (0,t)-3\varphi (0,t)\int_0^1 \varphi
(xs,t)\varphi (x(1-s),t)\ ds \\
&+2\int_0^1 \int_0^1 \varphi (x(1-s),t)\varphi
(xss',t)\varphi (xs(1-s'),t)ds'ds \\
&-\hat{q}(0,t)\varphi (x,t)-\hat{q}(x,t)\varphi (0,t)+2\int_0^1 \varphi (x(1-s),t)\hat{q}(xs,t)ds+ \hat{q}_t (x,t).
\end{split}
\]

In Appendix it is shown that the determining equation is reduced to the study the  equation
\begin{equation}\label{remain2}
(c_2t-c_3)\hat{q}_t-c_0x\hat{q}_x+(c_1x+2c_2)\hat{q}=0,
\end{equation}
where the admitted generator has the form
\[
X=c_0X_0+c_1X_1+c_2X_2+c_3X_3,
\]
with
\begin{equation}\label{generatorX}
X_0=x\partial _x,\quad X_1=x\varphi\partial _{\varphi},\quad X_2=\varphi\partial _{\varphi}-t\partial_t,  \quad X_3=\partial _t.
\end{equation}

The values of the constants $c_0, c_1,c_2,$ and $c_3$ and relations between them depend on the function $\hat{q}$.
The trivial case, where the function $\hat{q}=0$ was studied in \cite{bk:GrigMel[1987]}, where it shown that the admitted Lie algebra is four-dimensional spanned by the generators $X_0$, $X_1$, $X_2$ and $X_3$.

\subsection{On Equivalence Transformations}

Let us introduce the operator $L$:
\begin{equation}\label{Lu}
L\varphi = \varphi _t(x,t)+\varphi (x,t)\varphi (0,t)-\int_0^1\ \varphi
(xs,t)\varphi (x(1-s),t)\ ds.
\end{equation}
Considering the transformations of $L\varphi$ corresponding to the generators $X_0$, $X_1$, $X_2$ and $X_3$, some equivalence transformations can be obtained.

The transformations corresponding to the generator $X_0=x\partial_x$
map a function $\varphi (x,t)$ into the function $ \bar{\varphi} (\bar{x},\bar{t})=\varphi (\bar{x}e^{-a},\bar{t}), $ where $a$ is the group parameter. The transformed expression of (\ref{Lu}) becomes
\[
\begin{array}{cl}
\bar{L}\bar{\varphi}&=\bar{\varphi} _{\bar{t}}(\bar{x},\bar{t})+\bar{\varphi} (\bar{x},\bar{t})\bar{\varphi} (0,\bar{t})-\int_0^1\ \bar{\varphi}(\bar{x}s,\bar{t})\bar{\varphi} (\bar{x}(1-s),\bar{t})\ ds \\
&=\varphi _{\bar{t}}(\bar{x}e^{-a},\bar{t})+\varphi (\bar{x}e^{-a},\bar{t})\varphi (0,\bar{t})-\int_0^1\ \varphi(\bar{x}e^{-a}s,\bar{t})\varphi (\bar{x}e^{-a}(1-s),\bar{t})\ ds \\
&=\varphi _t(x,t)+\varphi (x,t)\varphi (0,t)-\int_0^1\ \varphi
(xs,t)\varphi (x(1-s),t)\ ds \\
&=L\varphi.
\end{array}
\]
This defines the Lie group of equivalence transformations of equation (\ref{BEs})
\[
\bar{x}=xe^a, \quad \bar{t}=t, \quad \bar{\varphi}=\varphi, \quad \bar{\hat{q}}=\hat{q}.
\]

Similarly, the transformations corresponding to the generator $X_3=\partial_t$ define the equivalence Lie group
\[
\bar{x}=x, \quad \bar{t}=t+a, \quad \bar{\varphi}=\varphi, \quad \bar{\hat{q}}=\hat{q}.
\]

The transformations corresponding to the generator $X_2=\varphi\partial _{\varphi}-t\partial_t$ map a function $\varphi (x,t)$ into the function
$
\bar{\varphi} (\bar{x},\bar{t})=e^a\varphi (\bar{x},\bar{t}e^a),
$
which gives
\[
\begin{array}{cl}
\bar{L}\bar{\varphi}&=\bar{\varphi} _{\bar{t}}(\bar{x},\bar{t})+\bar{\varphi} (\bar{x},\bar{t})\bar{\varphi} (0,\bar{t})-\int_0^1\ \bar{\varphi}(\bar{x}s,\bar{t})\bar{\varphi} (\bar{x}(1-s),\bar{t})\ ds \\
&=e^a\varphi _{\bar{t}}(\bar{x},\bar{t}e^a)+e^a\varphi (\bar{x},\bar{t}e^a)e^a\varphi (0,\bar{t}e^a)-\int_0^1\ e^{2a}\varphi(\bar{x}s,\bar{t}e^a)\varphi (\bar{x}(1-s),\bar{t}e^a)\ ds \\
&=e^{2a}\varphi _{t}(\bar{x},\bar{t}e^a)+e^{2a}\varphi (\bar{x},\bar{t}e^a)\varphi (0,\bar{t}e^a)-e^{2a}\int_0^1\ \varphi(\bar{x}s,\bar{t}e^a)\varphi (\bar{x}(1-s),\bar{t}e^a)\ ds \\
&=e^{2a}\big(\varphi _t(x,t)+\varphi (x,t)\varphi (0,t)-\int_0^1\ \varphi
(xs,t)\varphi (x(1-s),t)\ ds \big)\\
&=e^{2a}L\varphi.
\end{array}
\]
Hence one can conclude that the transformations
\[
\bar{x}=x, \quad \bar{t}=te^{-a}, \quad \bar{\varphi}=\varphi e^a, \quad \bar{\hat{q}}=\hat{q}e^{2a}
\]
compose an equivalence Lie group of equation (\ref{BEs}).

The transformations corresponding to the generator $X_1=x\varphi\partial _{\varphi}$ map a function $\varphi (x,t)$ into the function $ \bar{\varphi} (\bar{x},\bar{t})=e^{\bar{x}a}\varphi (\bar{x},\bar{t})$, thus
\[
\begin{array}{cl}
\bar{L}\bar{\varphi}&=\bar{\varphi} _{\bar{t}}(\bar{x},\bar{t})+\bar{\varphi} (\bar{x},\bar{t})\bar{\varphi} (0,\bar{t})-\int_0^1\ \bar{\varphi}(\bar{x}s,\bar{t})\bar{\varphi} (\bar{x}(1-s),\bar{t})\ ds \\
&=e^{\bar{x}a}\varphi _{\bar{t}}(\bar{x},\bar{t})+e^{\bar{x}a}\varphi (\bar{x},\bar{t})\varphi (0,\bar{t})-\int_0^1\ e^{\bar{x}as}\varphi(\bar{x}s,\bar{t})e^{\bar{x}(1-s)a}\varphi (\bar{x}(1-s),\bar{t})\ ds \\
&=e^{\bar{x}a}\varphi _{t}(\bar{x},\bar{t})+e^{\bar{x}a}\varphi (\bar{x},\bar{t})\varphi (0,\bar{t})-e^{\bar{x}a}\int_0^1\ \varphi(\bar{x}s,\bar{t})\varphi (\bar{x}(1-s),\bar{t})\ ds \\
&=e^{xa}\big(\varphi _t(x,t)+\varphi (x,t)\varphi (0,t)-\int_0^1\ \varphi
(xs,t)\varphi (x(1-s),t)\ ds \big)\\
&=e^{xa}L\varphi,
\end{array}
\]
which gives the equivalence Lie group of transformations
\[
\bar{x}=x, \quad \bar{t}=t, \quad \bar{\varphi}=\varphi e^{xa}, \quad \bar{\hat{q}}=\hat{q}e^{xa}.
\]

Thus it has been shown that the generators
\begin{equation}\label{equivalentBE}
X_0^e=x\partial_x,\quad X_1^e=x\varphi\partial_{\varphi}+x\hat{q}\partial_{\hat{q}},\quad X_2^e=\varphi \partial_{\varphi}-t\partial_t+x\hat{q}\partial_{\hat{q}},  \quad X_3^e=\partial_t.
\end{equation}
define an equivalence Lie group of equation (\ref{BEs}).

Notice also that the transformation
\begin{equation}\label{involution}
E: \quad \bar{t}=-t,\quad \bar{\varphi}=-\varphi.
\end{equation}
does not change equation (\ref{BEs}). This is an involution.

Let us study the change of a generator $X=x_0X_0+x_1X_1+x_2X_2+x_3X_3$ under the transformations corresponding to these equivalence transformations. After the change defined by an equivalence transformation one gets the generator
\begin{equation}\label{decomposition}
X=\hat{x}_{0}\hat{X}_{0}+\hat{x}_{1}\hat{X}_{1}+\hat{x}_{2}\hat{X}_{2}+\hat{x}_{3}\hat{X}_{3},
\end{equation}
 where
\[
\hat{X}_{0}=\bar{x}\partial_{\bar{x}},\quad\hat{X}_{1}=\bar{x}\bar{\varphi}\partial_{\bar{\varphi}},\quad
\hat{X}_{2}=\bar{\varphi}\partial_{\bar{\varphi}}-\bar{t}\partial_{\bar{t}},\quad\hat{X}_{3}=\partial_{\bar{t}}.
\]
The corresponding transformations of the basis generators are
\begin{align*}
X_{0}^{{\rm e}}\colon & \,\, X_{0}=\hat{X}_{0},X_{1}=e^{-a}\hat{X}_{1},X_{2}=\hat{X}_{2},X_{3}=\hat{X}_{3};\\
X_{1}^{{\rm e}}\colon & \,\, X_{0}=\hat{X}_{0}+a\hat{X}_{1},X_{1}=\hat{X}_{1},X_{2}=\hat{X}_{2},X_{3}=\hat{X}_{3};\\
X_{2}^{{\rm e}}\colon & \,\, X_{0}=\hat{X}_{0},X_{1}=\hat{X}_{1},X_{2}=\hat{X}_{2},X_{3}=e^{-a}\hat{X}_{3};\\
X_{3}^{{\rm e}}\colon & \,\, X_{0}=\hat{X}_{0},X_{1}=\hat{X}_{1},X_{2}=\hat{X}_{2}+a\hat{X}_{3},X_{3}=\hat{X}_{3}.
\end{align*}
or coordinates of the generator $X$ are changed as follows
\[
\begin{array}{cl}
X_{0}^{{\rm e}}\colon & \hat{x}_{1}=x_{1}e^{-a},\\
X_{1}^{{\rm e}}\colon & \hat{x}_{1}=x_{1}+ax_{0},\\
X_{2}^{{\rm e}}\colon & \hat{x}_{3}=x_{3}e^{-a},\\
X_{3}^{{\rm e}}\colon & \hat{x}_{3}=x_{3}+ax_{2},
\end{array}
\]
where only changeable coordinates are presented.

\subsection{Algebraic approach for analyzing equation (\ref{remain2})}

Group classification of equation (\ref{BEs}) is carried out up to the
equivalence transformations considered in the previous section.
The method for classifying the source function $\hat{q}$ is similar to
the method which was used for classifying equations for moment generating
function in
\cite{bk:GrigorievMeleshkoSuriyawichitseranee[2014]}.
First of all one notes that actions of equivalence transformations
corresponding to the equivalence Lie algebra spanned by the generators
(\ref{equivalentBE}) is equivalent to the automorphisms of the Lie algebra
$L_4$ spanned by the generators $X_{0},X_{1},X_{2},X_{3}$.

In fact, the commutators of these generators are
\[
\begin{array}{c|cccc}
 & X_{0} & X_{1} & X_{2} & X_{3}\\
\hline X_{0} & 0 & X_{1} & 0 & 0\\
X_{1} & -X_{1} & 0 & 0 & 0\\
X_{2} & 0 & 0 & 0 & -X_{3}\\
X_{3} & 0 & 0 & X_{3} & 0
\end{array}
\]
Using table of commutators, the inner automorphisms are obtained
\[
\begin{array}{rl}
A_{0}\colon & \hat{x}_{1}=x_{1}e^{a},\\
A_{1}\colon & \hat{x}_{1}=x_{1}+ax_{0},\\
A_{2}\colon & \hat{x}_{3}=x_{3}e^{a},\\
A_{3}\colon & \hat{x}_{3}=x_{3}+ax_{2},
\end{array}
\]
where only changeable coordinates are presented.

Thus one can conclude that the changes corresponding to the equivalence transformations are similar to the actions of the inner automorphisms. Because of this property one can use an optimal system of subalgebras for classifying equation (\ref{BEs}).

The commutator table of Lie algebra $L_4$ coincides with the commutator table considered in \cite{bk:GrigorievMeleshkoSuriyawichitseranee[2014]}, where group classification of the equation for a moment generating function was studied. The difference in constructing an optimal system here consists of the set of involutions: in the present case the involution corresponding to $\hat{x}_1=-x_1$ is absent comparing with \cite{bk:GrigorievMeleshkoSuriyawichitseranee[2014]}.
The optimal system of subalgebras of the Lie algebra $L_4$ is presented in Table 1,
where $\gamma$ is an arbitrary constant.
\begin{table}[!h]
\label{tab:1}
\begin{center}
\caption{Optimal system of subalgebras of $L_{4}$}
\end{center}
\begin{center}
\vspace{0 cm}
\begin{tabular}{ c l | c l}
\hline
No. & Basis  &No.  & Basis\\
\hline
1.  & $X_{0},\: X_{1},\: X_{2},\: X_{3}$      & 13.  & $X_{0}+X_{3},\: X_{1}$\\
2.  & $\gamma X_{0}+X_{2},\: X_{1},\: X_{3}$  & 14.  & $X_{1},\: X_{3}$\\

3.  & $X_{0},\: X_{1},\: X_{3}$               & 15.  & $X_{0},\: X_{3}$\\

4.  & $X_{0},\: X_{1},\: X_{2}$               & 16.  & $X_{0},\: X_{1}$\\

5.  & $X_{0},\: X_{2},\: X_{3}$               & 17.  & $\gamma X_{0}+ X_{2}$\\

6.  & $X_{2},\: X_{3}$                        & 18.  & $X_{1}+X_{2}$\\

7.  & $X_{2}+X_{0},\: X_{1}+\: X_{3}$         & 19.  & $X_{1}-X_{2}$\\

8.  & $X_{2}+\gamma X_{0},\: X_{3}$           & 20.   & $X_{0}+X_{3}$\\

9.  & $X_{1}+X_{2},\: X_{3}$                 & 21.   & $X_{1}+X_{3}$\\

10. & $X_{1}-X_{2},\: X_{3}$                 & 22.   & $X_{0}$\\

11. & $X_{0},\: X_{2}$                       & 23.   & $X_{1}$\\

12. & $\gamma X_{0}+X_{2},\: X_{1}$          & 24.  & $X_{3}$\\

\hline
\end{tabular}
\vspace{-.5 cm}
\end{center}
\end{table}

Using the optimal system of subalgebras for  group classification of equation (\ref{BEs}), the functions $\hat{q}(t,x)$ are obtained by substituting into equation (\ref{remain2}) the constants $c_{i}$ corresponding to the basis generators of a subalgebra of the optimal system of subalgebras,
and solving the obtained system of equations. The result of the group
classification is presented in Table 2, where $\beta$ and $\gamma$ are constant, and the function $\Phi$ is an arbitrary function of its argument.

\begin{table}[!h]
\vspace{0.1 cm}
\begin{center}
\caption{Group classification of equation (\ref{BEs})}
\end{center}
\begin{center}
\vspace{0 cm}
\begin{tabular}{ l l l }
\hline
No. & $\hat{q}(t,x)$  &\ \ \  Generators  \\
\hline
1.  & $0$  & $X_{0},\: X_{1},\: X_{2}\: X_{3}$   \\

2.  & $\beta x^{2}e^{tx}$  & $X_{2}+X_{0},\: X_{1}+ X_{3}$ \\

3.  & $\beta x{}^{\gamma}$  & $\gamma X_{2}+2X_{0},\: X_{3}$  \\

4.  & $\beta t{}^{-2}$  & $X_{0},\: X_{2}$       \\

5.  & $t^{-2}\Phi(xt^{\gamma})$  & $\gamma X_0+X_2$  \\

6.  & $t^{-(x+2)}\Phi(x)$  & $X_1+X_2$ \\

7.  & $t^{x-2}\Phi(x)$  & $X_1-X_2$ \\

8.  & $\Phi(xe^{-t})$  & $X_{0}+X_{3}$ \\

9. & $e^{xt}\Phi(x)$   & $X_{1}+X_{3}$ \\

10. & $\Phi(t)$  & $X_{0}$ \\

11. & $\Phi(x)$  & $X_{3}$ \\
\hline

\end{tabular}
\vspace{-.5 cm}
\end{center}
\end{table}

\subsection{Illustrative Examples of Obtaining the Function $\hat{q}$}

\subsubsection{Lie algebra $\{ \gamma X_2+2 X_0,\, X_3 \}$}

This case corresponds to the case $3$ in Table 2. For this Lie algebra, there are two sets of the coefficients $c_i$, $(i=0,1,2,3)$:
\[
    \begin{array}{rllll}
     \gamma X_2+2 X_0: &c_0=2 & c_1=0 & c_2=\gamma & c_3=0; \\
    X_3: &c_0=0 & c_1=0 & c_2=0 & c_3=1.
   \end{array}
\]
These sets define the system of equations by substituting the coefficients $c_i$ into equation (\ref{remain2}):
\[
\gamma(\frac 12 t\hat{q}_t+\hat{q})-x\hat{q}_x =0, \qquad \hat{q}_t =0.
\]
The general solution of these equations is $\hat{q}=\beta x^\gamma$, where $\beta$ is constant.

\subsubsection{Lie algebra $\{ X_1-X_2\}$}

For this algebra there is only one equation
\[
t\hat{q}_t+2\hat{q}- x\hat{q}_x =0.
\]
The general solution of this equation is $\hat{q}=t^{x-2}\Phi (x) $, where $\Phi(x)$ is an arbitrary function. This case corresponds to the case $7$ in Table 2.

\section{Invariant Solutions}

The study presented in this section is devoted to constructing invariant solutions of equation (\ref{BEs}). For each obtained function $\hat{q}$, we study the admitted Lie algebra. Using an optimal system of subalgebras of these Lie algebras, we derive invariant solutions. The set of all these solutions defines the set of all possible invariant solutions of equation (\ref{BEs}). It is shown here that similar to differential equations, equations for finding invariant solutions are reduced to equations with fewer number of the independent variables.

\subsection{Invariant Solutions of equation (\ref{BEs}) with $\hat{q}=\beta x^{2}e^{xt}$ }

For the source function $\hat{q}=\beta x^{2}e^{tx}$ the admitted Lie algebra of equation (\ref{BEs}) is $\{X_{2}+X_{0},\: X_{1}+ X_{3}\}$. An optimal system of subalgebras of this Lie algebra consists of the subalgebras:
\[
\{X_{2}+X_{0}\},\ \  \{X_{1}+ X_{3}\},\ \  \{X_{2}+X_{0},\, X_{1}+ X_{3}\}.
\]

A representation of an invariant solution corresponding to the subalgebra $\{X_{2}+X_{0}\}$ is $\varphi = t^{-1}r(z)$, where $z=xt$. Substituting this representation of a solution into equation (\ref{BEs}), it becomes
\begin{equation}\label{invariant1}
zr'(z)-r(z)+r(z)r(0)-\int_0^1\ r(zs)r(z(1-s))\ ds = \beta z^2e^{z}.
\end{equation}
The latter equation (\ref{invariant1}) is an equation with the single independent variable $z$

A representation of an invariant solution corresponding to the subalgebra $\{X_{1}+X_{3}\}$ is $\varphi = e^{xt}r(x)$. Substituting this representation of a solution into equation (\ref{BEs}), the reduced equation is obtained
\[
r(x)(x+r(0))-\int_0^1\ r(xs)r(x(1-s))\ ds = \beta x^2.
\]

The subalgebra $\{X_{2}+X_{0},\: X_{1}+ X_{3}\}$ gives an invariant solution in the form $\varphi = Cxe^{xt}$, where $C$ is constant. After substituting this representation into equation (\ref{BEs}), we get the equation for the constant $C$,
\[
C^2-6C+6\beta = 0.
\]
If $\beta \leq \frac{3}{2}$, then $C = 3 \pm \sqrt{9-6\beta}$.

\subsection{Invariant Solutions of equation (\ref{BEs}) with $\hat{q}=\beta x^{\gamma}$ }
An optimal system of subalgebras consists of the list: $\{ \gamma X{}_{2}+2X{}_{0},\:X{}_{3} \}$, $\{ X{}_{3}\}$ and  either $\{ \gamma X{}_{2}+2X{}_{0} \}$ for $\gamma \neq 0$ or $\{X{}_{3}+\alpha X{}_{0}\}$ for $\gamma = 0$.

A representation of an invariant solution corresponding to the subalgebra $\{ \gamma X{}_{2}+2X{}_{0},\: X{}_{3} \}$ is $\varphi=Cx^{\frac{\gamma}{2}}$, where $C$ is constant. After substituting this representation into equation (\ref{BEs}), we get the equation for the constant $C$,
\[ C^2B+\beta=0, \] where $B=\intop_{0}^{1}(s(1-s))^{\frac{\gamma}{2}}ds$.

A representation of an invariant solution corresponding to the subalgebra $\{ X{}_{3}\}$ is $\varphi=r(x)$.
Substituting this representation into equation (\ref{BEs}), the reduced equation is
\[
r(x)r(0)-\int_{0}^{1}r(xs)r(x(1-s))ds=\beta x^{\gamma}.
\]

A representation of an invariant solution corresponding to the subalgebra $\{ \gamma X{}_{2}+2X{}_{0} \}$ where $\gamma \neq 0$ is $\varphi=t^{-1}r(z)$, where $z=t{}^{2}x^{\gamma}$.
Substituting this representation into equation (\ref{BEs}), one has the reduced equation
\[
2zr'(z)-r(z)+r(z)r(0)-\int_{0}^{1}r(zs)r(z(1-s))ds=\beta z
\]

A representation of an invariant solution corresponding to the subalgebra $\{X_{3}+\alpha X_{0}\}$  with $\gamma = 0$ is $\varphi=r(z)$, where $z=xe^{-\alpha t}$.
Substituting this representation into equation (\ref{BEs}), one has
\[
-\alpha zr'(z)+r(z)r(0)-\int_{0}^{1}r(zs)r(z(1-s))ds=\beta .
\]

\subsection{Invariant Solutions of equation (\ref{BEs}) with $\hat{q}=\beta t^{-2}$ }
For the source function $\hat{q}=\beta t^{-2}$ the admitted Lie algebra of equation (\ref{BEs}) is $\{X_{0},\:X_{2}\}$. An optimal system of subalgebras of this Lie algebra consists of the subalgebras:
$\{ X{}_{0},\:X_{2} \}$, $\{ X{}_{2}+\alpha X_{0}\}$ and $\{X_0\}$.

An invariant solution corresponding to the subalgebra $\{ X{}_{0},\:X_{2} \}$ is $\varphi=-\beta t^{-1}$.

A representation of an invariant solution corresponding to the subalgebra
$\{ X_{2}+\alpha X_{0}\}$ is $\varphi=t^{-1}r(z)$, where $z=xt^{\alpha}$.
Substituting this representation into equation (\ref{BEs}), one obtains the equation
\[
\alpha zr'(z)-r(z)+r(z)r(0)-\int_{0}^{1}r(zs)r(z(1-s))ds=\beta
.
\]

A representation of an invariant solution corresponding to the subalgebra $\{X_0\}$  is $\varphi=r(t)$, which gives the the solution $\varphi=-\beta t^{-1}+C$, where $C$ is constant.

\subsection{Invariant Solutions of equation (\ref{BEs}) with $\hat{q}=t^{-2}\Phi(xt^{\gamma })$ }
In this case the admitted Lie algebra is $\{ \gamma  X_{0}+X_{2}\} $. An invariant solution has the representation $\varphi=t^{-1}r(z)$, where $z=xt{}^{\gamma }$. Substituting this representation into equation (\ref{BEs}), the reduced equation is
\[ \gamma  zr'(z)-r(z)+r(z)r(0)-\int_{0}^{1}r(zs)r(z(1-s))ds=\Phi(z). \]

\subsection{Invariant Solutions of equation (\ref{BEs}) with $\hat{q}=t^{-(x+2)}\Phi(x)$ }
The admitted Lie algebra is $\{X_{1}+X_{2}\}$. An invariant solution has the representation $\varphi=t^{-(x+1)}r(x)$. Substituting this representation into equation (\ref{BEs}), the reduced equation is
\[ -(x+1)r(x)+r(x)r(0)-\int_{0}^{1}r(xs)r(x(1-s))ds=\Phi(x). \]

\subsection{Invariant Solutions of equation (\ref{BEs}) with $\hat{q}=t^{x-2}\Phi(x)$ }
The admitted Lie algebra is $\{X_{1}-X_{2}\}$. An invariant solution has the representation $\varphi=t^{x-1}r(x)$. The reduced equation is
\[ (x-1)r(x)+r(x)r(0)-\int_{0}^{1}r(xs)r(x(1-s))ds=\Phi(x). \]

\subsection{Invariant Solutions of equation (\ref{BEs}) with $\hat{q}=\Phi(xe^{-t})$ }
For the function $\hat{q} = \Phi(xe^{-t})$, the admitted Lie algebra is $\{X_0+X_3\}$. Its invariant solution has the representation $ \varphi =r(z) $, where $z= xe^{-t}$.
Substituting this representation into equation (\ref{BEs}), the reduced equation is
\[
-zr'(z)+r(z)r(0)-\int_0^1\ r(zs)r(z(1-s))\ ds = \Phi(z).
\]
In particular, for BKW-solution $r=6e^z(1-z)$ which gives that $\Phi =0$.

\subsection{Invariant Solutions of equation (\ref{BEs}) with $\hat{q}=e^{xt}\Phi(x)$ }
The admitted Lie algebra is $\{X_1+X_3\}$. An invariant solution has the representation $\varphi=e^{xt}r(x) $. Substituting this representation into equation (\ref{BEs}), one obtains
\[
xr(x)+r(x)r(0)-\int_{0}^{1}r(xs)r(x(1-s))ds=\Phi(x).
\]

\subsection{Invariant Solutions of equation (\ref{BEs}) with $\hat{q}=\Phi(t)$ }
The admitted Lie algebra is $\{X_{0}\}$. Its invariant solution is
$\varphi=\int\Phi(t)\: dt$.

\subsection{Invariant Solutions of equation (\ref{BEs}) with $\hat{q}=\Phi(x)$ }
The admitted Lie algebra is $\{X_3\}$. An invariant solution has the representation $\varphi=r(x) $.
Substituting this representation into equation (\ref{BEs}), the reduced equation is
\[
r(x)r(0)-\int_{0}^{1}r(xs)r(x(1-s))ds=\Phi(x).
\]

\section{Acknowledgement}

The authors are thankful to E.Schulz for valuable remarks.

\section{Appendix}

\subsection{Solving the determining equation}

Assuming  that the coefficients of the infinitesimal generator $X$ are represented by the formal Taylor series with respect to $\varphi$, one has that
\begin{equation}\label{cogen}
\begin{split}
\xi (x,t,\varphi)=\sum\limits_{l \geq 0} q_l(x,t) \varphi^l(x,t),\\
\eta (x,t,\varphi)=\sum\limits_{l \geq 0} r_l(x,t) \varphi^l(x,t), \\
\zeta (x,t,\varphi)=\sum\limits_{l \geq 0} p_l(x,t) \varphi^l(x,t).
\end{split}
\end{equation}

For solving the determining equation we use a particular class of solutions
corresponding to the initial data
\begin{equation}\label{initial}
\varphi_0(x) = bx^n
\end{equation}
at a given (arbitrary) time $t=t_0$. Here $n$ is a positive integer and $b$ is a real number. Notice that the coefficients of the admitted generator do not depend on solution. Varying the parameter $b$ and the degree $n$ in (\ref{initial}), one can obtain conditions for the coefficients.

In the case of $n=0$ the derivatives of the function $\varphi(x,t)$ become
\[
\varphi _t(x,t)= \hat{q} (x,t), \qquad \varphi _{xt}(x,t)= \hat{q}_x (x,t),
\]
\[
\varphi _{tt}(x,t)=-\hat{q}(0,t)b-\hat{q}(x,t)b+2\int_0^1 b \hat{q}(xs,t)ds+ \hat{q}_t (x,t).
\]
The determining equation becomes
\begin{equation}
\begin{array}{cc}
&\zeta_t(x,t,b)+\hat{q}(x,t) \zeta_{\varphi}(x,t,b)-\hat{q}_x (x,t) \xi(x,t,b)-\eta_t (x,t,b) \hat{q}(x,t)-\hat{q}^{2} (x,t) \eta_{\varphi}(x,t,b)\\
&-\eta(x,t,b)(-\hat{q}(0,t)b-\hat{q}(x,t)b+2\int_0^1 b \hat{q}(xs,t)ds+ \hat{q}_t (x,t)) \\
&+(\zeta(0,t,b)-\eta(0,t,b)\hat{q}(0,t))b+(\zeta(x,t,b)-\eta(x,t,b) \hat{q}(x,t))b \\
&-2\int_0^1 b(\zeta(xs,t,b) - \eta(xs,t,b) \hat{q}(xs,t))ds =0.
\end{array}
\end{equation}
Using the decompositions (\ref{cogen}), from this equation one obtains
\begin{equation}
\label{n01}
\begin{array}{c}
\frac{\partial p_0(x,t)}{\partial t} + \hat{q}(x,t)p_1(x,t)-\hat{q}_x(x,t)q_0(x,t)
-\hat{q}(x,t)\frac{\partial r_0}{\partial t}\\
-\hat{q}^2(x,t)r_1(x,t)-\hat{q}_t(x,t) r_0(x,t) =0
\end{array}
\end{equation}
and
\begin{equation}\label{n02}
\begin{array}{cc}
&\frac{\partial p_{l+1}(x,t)}{\partial t} +(l+2)\hat{q}(x,t)p_{l+2}(x,t)-\hat{q}_x(x,t)q_{l+1}(x,t)-\hat{q}(x,t)\frac{\partial r_{l+1}(x,t)}{\partial t} \\
&-(l+2)\hat{q}^2(x,t)r_{l+2}(x,t)+\hat{q}(0,t)r_l(x,t)-2r_l(x,t)\int_0^1 \hat{q}(xs,t)ds \\
&-\hat{q}_t(x,t) r_{l+1}(x,t)+p_l(0,t)-\hat{q}(0,t) r_l(0,t)+p_l(x,t)-2\int_0^1 p_l(xs,t)ds \\
&+ 2\int_0^1 r_l(xs,t)\hat{q}(xs,t)ds =0
\end{array}
\end{equation}
where $l=0,1,2,3,....$

If $n \geq 1$ the derivatives of the function $\varphi(x,t)$ at the time $t=t_0$ are
\[
\begin{split}
\varphi _t(x,t)= b^2 x^{2n}P_n +\hat{q} (x,t),\quad \varphi _{xt}(x,t)= 2n b^2 x^{2n-1} P_n +\hat{q}_x (x,t),\\
\varphi _{tt}(x,t)=b^3 x^{3n} Q_n -\hat{q}(0,t)bx^n+2bx^n \int_0^1 (1-s)^n \hat{q}(xs,t)ds+ \hat{q}_t (x,t).
\end{split}
\]
Here the notations
\[
P_n=\frac{(n!)^2}{(2n+1)!},\quad Q_n=2P_n \frac{(2n)!n!}{(3n+1)!}
\]
are used. For further calculations we also notice that
\[
B(m+1,n+1)= \int_0^1 s^m(1-s)^n ds = \frac{m!n!}{(m+n+1)!}.
\]
In this case the determining equation (\ref{DEeq}) becomes
\begin{equation}\label{DEbn}
\begin{split}
\zeta_t&+\hat{q} \zeta_{\varphi}-\hat{q}_x \xi - \hat{q} \eta_t -\hat{q}^{2} \eta_{\varphi}- \hat{q}_t \eta  \\
+b &\bigg( -nx^{n-1} \xi_t - nx^{n-1} \hat{q} \xi_{\varphi} + x^n \hat{q} (0,t) \eta -2x^n\eta \int_0^1 (1-s)^n \hat{q}(xs,t)ds   \\
& +x^n \zeta (0,t,0) -2x^n \int_0^1 (1-s)^n\zeta(xs,t,b(xs)^n)ds \\
& -x^n \hat{q}(0,t) \eta(0,t,0)+2x^n \int_0^1 (1-s)^n \eta(xs,t,b(xs)^n)\hat{q}(xs,t)ds \bigg) \\
 +b^2 &\bigg( x^{2n} P_n \zeta_{\varphi}- 2nx^{2n-1} P_n \xi -x^{2n} P_n \eta_t-2x^{2n} P_n \hat{q}   \eta_{\varphi} -\delta_{n1} x^{2n-1} \xi(0,t,0) \\
&+2nx^{2n-1} \int_0^1 (1-s)^n s^{n-1} \xi(xs,t,b(xs)^n) ds \bigg)  \\
 +b^3 &\bigg( -nP_nx^{3n-1} \xi_{\varphi} - Q_nx^{3n}\eta + 2P_nx^{3n} \int_0^1 (1-s)^n s^{2n} \eta(xs,t,b(xs)^n) ds \bigg)  \\
+b^4 &\bigg( -P^2_n x^{4n} \eta_ {\varphi} \bigg)=0.
\end{split}
\end{equation}
Using the arbitrariness of the value $b$ and equating to zero the coefficients with respect to $b^\alpha \:(\alpha=0, 1, 2, ...),$ the determining equation can be split into a series of equations.

For $\alpha=0$ the corresponding coefficient vanishes due to equation (\ref{n01}).
For $\alpha=1$ the corresponding coefficient is
\[
\begin{split}
x^n \bigg(& \frac{\partial p_1(x,t)}{\partial t} +2p_2(x,t) \hat{q}(x,t) -\hat{q}_x(x,t) q_1(x,t) -\hat{q} (x,t) \frac{\partial r_1(x,t)}{\partial t}\\
&-2\hat{q}^2(x,t)r_2(x,t)-\hat{q}_t(x,t)r_1(x,t)+\hat{q}(0,t)r_0(x,t)+p_0(0,t)\\
&-\hat{q}(0,t)r_0(0,t)-2r_0(x,t)\int_0^1 (1-s)^n \hat{q}(xs,t)ds \\
&- 2\int_0^1 (1-s)^n p_0(xs,t)ds+2\int_0^1 (1-s)^n \hat{q}(xs,t)r_0(xs,t)ds \bigg)\\
+x^{n-1}& \bigg( -n\frac{\partial q_0(x,t)}{\partial t}-n\hat{q}(x,t) q_1(x,t) \bigg)  =0.
\end{split}
\]
After substituting $\frac{\partial p_1}{\partial t}$ found from equation (\ref{n02}) the latter equation is reduced to the equation
\begin{equation}\label{b11}
\begin{split}
x \bigg(& 2r_0(x,t)\int_0^1 \hat{q}(sx,t)ds  - p_0(x,t) + 2\int_0^1 p_0(xs,t) ds \\
&- 2\int_0^1 r_0(xs,t)\hat{q}(xs,t)ds -2r_0(x,t)\int_0^1 (1-s)^n \hat{q}(xs,t)ds  \\
&+2\int_0^1 (1-s)^n \hat{q}(xs,t)r_0(xs,t)ds - 2\int_0^1 (1-s)^n p_0(xs,t)ds\bigg)\\
 -\bigg( & n\frac{\partial q_0(x,t)}{\partial t}+n\hat{q}(x,t) q_1(x,t) \bigg)  =0.
\end{split}
\end{equation}
Dividing by $n$ equation (\ref{b11}), and considering  $n$ approaching to infinity, one obtains
\begin{equation}\label{b12}
\frac{\partial q_0}{\partial t}+ \hat{q} q_1   =0.
\end{equation}
Substituting (\ref{b12}) into (\ref{b11}) and approaching $n$ to infinity again, one gets
\begin{equation}\label{b14}
\begin{split}
 &2r_0(x,t)\int_0^1 \hat{q}(sx,t)ds  - p_0(x,t) + 2\int_0^1 p_0(xs,t) ds \\
 &- 2\int_0^1 r_0(xs,t)\hat{q}(xs,t) ds =0.
\end{split}
\end{equation}
Thus, equation (\ref{b11}) is reduced to the equation
\begin{equation}\label{b15}
\begin{split}
&r_0(x,t)\int_0^1 (1-s)^n\hat{q}(sx,t)ds  - \int_0^1 (1-s)^n r_0(xs,t)\hat{q}(xs,t) ds\\
&+ \int_0^1 (1-s)^n p_0(xs,t) ds =0.
\end{split}
\end{equation}
For analysing equation (\ref{b15}) one uses representations of the functions $\hat{q}, r_0$ and $p_0$ in the formal Taylor series:
\begin{equation}\label{rq0}
\hat{q} (x,t)= \sum\limits_{i=0}^{\infty}h_i(t)x^i, \qquad
r_0 (x,t)= \sum\limits_{i=0}^{\infty}r_{0i}(t)x^i, \qquad
p_0 (x,t)= \sum\limits_{i=0}^{\infty}p_{0i}(t)x^i.
\end{equation}
All coefficients with respect to $x^k$ in equation (\ref{b15}) have to vanish
\[
\sum\limits_{i=0}^k \big[ r_{0i}h_{k-i} \big( \frac{(k-i)!(n+k+1)!}{k!(n+k-i+1)!}-1 \big) \big] +p_{0k} = 0.
\]
Due to arbitrariness of $n$, one finds
\begin{equation}\label{b16}
 r_{0i}h_{k-i} = 0,\qquad p_{0k} = 0
\end{equation}
for all $i \geq 1$ and  $k \geq 0$. Since $\hat{q} \neq 0$, then there exists $k_0$ such that $h_{k_0} \neq 0$. Choosing $k = i+k_0$ for any $i \geq 1$, one gets from the first equation of (\ref{b16}) that
\begin{equation}\label{r0}
 r_0(x,t) = r_{00}(t).
\end{equation}
The second condition of (\ref{b16}) provides that
\begin{equation}\label{p0}
p_0(x,t)=0.
\end{equation}


For $\alpha=2$ the corresponding equation,
after substituting in it the expression for $\frac{\partial p_2}{\partial t}$ found from equation (\ref{n02}), becomes
\begin{equation}\label{b21}
\begin{split}
x \bigg(& - p_1(x,t)- p_1(0,t)+ 2\int_0^1 p_1(xs,t) ds+2r_1(x,t)\int_0^1 \hat{q}(sx,t)ds  +\hat{q}(0,t)r_1(0,t)   \\
&-2\int_0^1 r_1(xs,t)\hat{q}(xs,t)ds-2r_1(x,t)\int_0^1 (1-s)^n \hat{q}(xs,t)ds  \\
&-2\int_0^1 (1-s)^n s^np_1(xs,t) ds +2\int_0^1 (1-s)^n s^n \hat{q}(xs,t)r_1(xs,t)ds\\
&+P_n p_1(x,t)-P_n \frac{\partial r_0(x,t)}{\partial t}-2P_n \hat{q}(x,t) r_1(x,t) \bigg) \\
 +\bigg(&  -n\frac{\partial q_1(x,t)}{\partial t}-2n\hat{q}(x,t) q_2(x,t)+2n\int_0^1(1-s)^n s^{n-1} q_0(xs,t)ds\\
 &-2nP_nq_0(x,t)-\delta_{n1}q_0(0,t) \bigg)  =0.
\end{split}
\end{equation}
Dividing by $n$ and approaching $n$ to infinity, one has
\begin{equation}\label{b22}
\frac{\partial q_1}{\partial t}+ 2\hat{q} q_2   =0.
\end{equation}
Approaching $n$ to infinity in (\ref{b21}), one obtains
\begin{equation}\label{b24}
\begin{split}
x \bigg(& - p_1(x,t)- p_1(0,t)+ 2\int_0^1 p_1(xs,t) ds+2r_1(x,t)\int_0^1 \hat{q}(sx,t)ds  +\hat{q}(0,t)r_1(0,t)   \\
&-2\int_0^1 \hat{q}(xs,t)r_1(xs,t)ds \bigg) -\delta_{n1}q_0(0,t) =0.
\end{split}
\end{equation}
Equation (\ref{b21}) becomes
\begin{equation}\label{b25}
\begin{split}
x \bigg(&-2r_1(x,t)\int_0^1 (1-s)^n \hat{q}(xs,t)ds  +2\int_0^1 (1-s)^n s^n \hat{q}(xs,t)r_1(xs,t)ds \\
&-2\int_0^1 (1-s)^n s^np_1(xs,t) ds+P_n p_1(x,t)-P_n \frac{\partial r_0(x,t)}{\partial t}-2P_n \hat{q}(x,t) r_1(x,t) \bigg) \\
 +\bigg(&2n\int_0^1(1-s)^n s^{n-1} q_0(xs,t)ds-2nP_nq_0(x,t) \bigg)  =0.
\end{split}
\end{equation}
Using representations (\ref{rq0}) and
\[
r_1 (x,t)= \sum\limits_{i=0}^{\infty}r_{1i}(t)x^i,\qquad p_1 (x,t)= \sum\limits_{i=0}^{\infty}p_{1i}(t)x^i,
\]
equation (\ref{b25}) is reduced to the equation
\begin{equation} \label{b26}
\begin{split}
\sum\limits_{k=0}^{\infty} x^{k+1}\bigg[& \sum\limits_{i=0}^{k} r_{1i} h_{k-i} \big(
\frac{n!(k-i)!}{(n+k-i+1)!}-\frac{n!(n+k)!}{(2n+k+1)!}+\frac{n!n!}{(2n+1)!} \big) \\
&+p_{1k} \big(\frac{n!(n+k)!}{(2n+k+1)!}-\frac{n!n!}{2(2n+1)!} \big) \bigg] +x\frac{n!n!}{2(2n+1)!}\frac{\partial r_{00}}{\partial t} \\
+\sum\limits_{k=0}^{\infty} & x^k q_{0k} \big(\frac{nn!n!}{(2n+1)!}-\frac{nn!(n+k-1)!}{(2n+k)!}   \big) =0.
\end{split}
\end{equation}
Equating coefficients with respect to $x^0$ and $x^1$ to zero, one finds
\begin{equation}\label{q00}
q_{00} = 0,
\end{equation}
and
\[
r_{10}h_0 +\frac{n!(n+1)!}{2(2n+1)!} \big( p_{01}+ \frac{\partial r_{00}}{\partial t} \big) = 0.
\]
Due to arbitrariness of $n$ these equations provide that
\[
 r_{10}h_{0} = 0,
 \]
and
\begin{equation}\label{p10r}
p_{10}+\frac{\partial r_{00}}{\partial t} = 0.
\end{equation}
Equating the coefficients with respect to $x^k$ in (\ref{b26}) for $k\geq 2$, one have
\[
\begin{split}
&\sum\limits_{i=0}^{k-1}\bigg[ r_{1i} h_{k-1-i} \bigg(
\frac{(k-1-i)!}{(n+k-i)!}-\frac{(n+k-1)!}{(2n+k)!}+\frac{n!}{(2n+1)!} \bigg) \bigg] \\
&+p_{1(k-1)} \bigg(\frac{(n+k-1)!}{(2n+k)!}-\frac{n!}{2(2n+1)!} \bigg)\\
&+q_{0k}\bigg( \frac{n(n!)}{(2n+1)!}-\frac{n((n+k-1)!)}{(2n+k)!} \bigg)=0,
\end{split}
\]
which gives that
\[
r_{1i}h_{k-1-i}=0,\qquad q_{0k}=0, \qquad p_{1k}=0,
\]
for all $i=0,1,2, ...,k-1$ and $k=2,3,4,...$.
From the obtained conditions, one can conclude that
\begin{equation} \label{r1q}
r_1(x,t)\hat{q}(x,t)=0,
\end{equation}
\begin{equation}\label{q0}
q_0(x,t)=q_{01}(t)x,
\end{equation}
\begin{equation}\label{p1}
p_1(x,t)=p_{10}(t)+p_{11}(t)x.
\end{equation}
Since $\hat{q} \neq 0$, equation (\ref{r1q}) gives that $r_1 = 0$.


Returning to equation (\ref{DEbn}), for $\alpha=3$ the corresponding equation
after substituting into it $\frac{\partial p_3}{\partial t}$, found from equation (\ref{n02}), becomes
\begin{equation}\label{b31}
\begin{split}
x \bigg(& - p_2(x,t)- p_2(0,t)+ 2\int_0^1 p_2(xs,t) ds+2r_2(x,t)\int_0^1 \hat{q}(sx,t)ds  +\hat{q}(0,t)r_2(0,t)   \\
&-2\int_0^1 r_2(xs,t)\hat{q}(xs,t)ds-2r_2(x,t)\int_0^1 (1-s)^n \hat{q}(xs,t)ds  \\
&-2\int_0^1 (1-s)^n s^{2n}p_2(xs,t) ds +2\int_0^1 (1-s)^n s^{2n} \hat{q}(xs,t)r_2(xs,t)ds\\
&+2P_n p_2(x,t)-P_n \frac{\partial r_1(x,t)}{\partial t}-4P_n \hat{q}(x,t) r_2(x,t)-Q_nr_0(x,t) \\
&+2P_n\int_0^1 (1-s)^n s^{2n}r_0(xs,t)ds \bigg) \\
 +\bigg(&  -n\frac{\partial q_2(x,t)}{\partial t}-3n\hat{q}(x,t) q_3(x,t)-2nP_nq_1(x,t)\\
&+2n\int_0^1(1-s)^n s^{2n-1} q_1(xs,t)ds-nP_nq_1(x,t) \bigg)=0.
\end{split}
\end{equation}
Similarly, dividing by $n$ the latter equation, approaching $n$ to infinity and using the representations of the functions $\hat{q}, r_0, r_1, r_2, q_1$ and $p_2$ in the formal Taylor series, one obtains
\begin{equation}\label{b32}
\frac{\partial q_2}{\partial t}+ 3\hat{q} q_3   =0,
\end{equation}
\begin{equation}\label{b33}
q_1(x,t) = 0,\quad\frac{\partial r_1(x,t)}{\partial t}=0, \quad p_2(x,t)=0,\quad r_2(x,t)\hat{q}(x,t)=0.
\end{equation}
The last equation of (\ref{b33}) gives that $r_2 = 0$. From equations (\ref{b12}), (\ref{q0}) and the first condition of (\ref{b33}), one has
\begin{equation}
q_0(x,t)=c_0x.
\end{equation}

For $\alpha=4+l \;  (l \geq 0)$ equation (\ref{DEbn}) is reduced to the equation
\begin{equation}\label{b41}
\begin{split}
x \bigg(& - p_{3+l}(x,t)- p_{3+l}(0,t)+ 2\int_0^1 p_{3+l}(xs,t) ds+2r_{3+l}(x,t)\int_0^1 \hat{q}(sx,t)ds  +\hat{q}(0,t)r_{3+l}(0,t)   \\
&-2\int_0^1 r_{3+l}(xs,t)\hat{q}(xs,t)ds-2r_{3+l}(x,t)\int_0^1 (1-s)^n \hat{q}(xs,t)ds  \\
&-2\int_0^1 (1-s)^n s^{3n}p_{3+l}(xs,t) ds +2\int_0^1 (1-s)^n s^{3n} \hat{q}(xs,t)r_{3+l}(xs,t)ds\\
&+(3+l)P_n p_{3+l}(x,t)-P_n \frac{\partial r_{2+l}(x,t)}{\partial t}-2(3+l)P_n \hat{q}(x,t) r_{3+l}(x,t)-Q_nr_{1+l}(x,t) \\
&+2P_n\int_0^1 (1-s)^n s^{3n}r_{1+l}(xs,t)ds-(l+1)P_n^2 r_{1+l}(x,t) \bigg) \\
 +\bigg(& -n\frac{\partial q_{3+l}(x,t)}{\partial t}-(4+l)n\hat{q}(x,t) q_{4+l}(x,t)-2nP_nq_{2+l}(x,t))-(2+l)nP_nq_{2+l}(x,t)\\
&+2n\int_0^1(1-s)^n s^{(3+l)n-1} q_{2+l}(xs,t)ds \bigg)=0.
\end{split}
\end{equation}
Here $\frac{\partial p_{4+l}}{\partial t}$ found from equation (\ref{n02}) is also substituted.

Similar to the previous case, dividing by $n$ the latter equation, approaching $n$ to infinity, and using the representations of the functions, one obtains
\[
\begin{split}
\frac{\partial q_{3+l}(x,t)}{\partial t}+(4+l)\hat{q}(x,t) q_{4+l}(x,t)=0, \quad
q_{2+l}(x,t) = 0, \frac{\partial r_{2+l}(x,t)}{\partial t}=0,\\
 \quad p_{3+l}(x,t)=0,\quad r_{3+l}(x,t)\hat{q}=0, \quad r_{1+l}(x,t)=0,\quad (l \geq 0).
\end{split}
\]


Thus,
\[
\begin{split}
p_0(x,t) = 0, \quad p_1(x,t)=p_{10}(t)+p_{11}(t)x, \quad q_0(x,t)=c_0x \\
\frac{\partial r_k(x,t)}{\partial x}=0, \quad p_{k+1}=0, \quad q_k =0, \quad r_k =0, \quad k \geq 1.
\end{split}
\]

Substituting $p_0, p_2, q_1, r_1, r_2$ and $r_0(x,t)=r_{00}(t)$ into equation (\ref{n02}) in case $l=0$, one finds
\begin{equation}\label{dp1}
\frac{\partial p_{10}}{\partial t}=0, \quad \frac{\partial p_{11}}{\partial t}=0
\end{equation}
or $p_{10}(t)=c_2$, $p_{11}(t)=c_1$,  where $c_1$ and $c_2$ are constant.
Hence
\begin{equation}
p_1(x,t)=c_2+c_1x,
\end{equation}

Equation (\ref{p10r}) gives that
\begin{equation}
r_{00}(t)=-c_2t+c_3
\end{equation}
where $c_3$ is constant.

Therefore, the coefficients of the generator $X$ are
\begin{equation}\label{gen}
\xi (x,t,\varphi)=c_0x,\quad
\eta (x,t,\varphi)=-c_2t+c_3, \quad
\zeta (x,t,\varphi)=(c_2+c_1x)\varphi
\end{equation}
where $c_0, c_1, c_2 $ and $c_3$ are arbitrary constants.

\section*{References}


\begin{thebibliography}{10}

\bibitem{bk:BoffiSpiga[1982a]}
V.~C. Boffi and G.~Spiga.
\newblock Nonlinear diffusion of test particles in the presence of an external
  conservative force.
\newblock {\em Phys. Fluids}, 25:1987--1992, 1982.

\bibitem{bk:BoffiSpiga[1982c]}
V.~C. Boffi and G.~Spiga.
\newblock Global solution to a nonlinear integral evolution problem in particle
  transport theory.
\newblock {\em J. Math. Phys.}, 23(11):2299--2303, 1982.

\bibitem{bk:Nonnenmacher[1984]}
T.~F. Nonenmacher.
\newblock Application of the similarity method to the nonlinear {B}oltzmann
  equation.
\newblock {\em J. of Appl. Mathem. and Physics (ZAMP)}, 35(5):680--691, 1984.

\bibitem{bk:KW}
M.~Krook and T.~T. Wu.
\newblock Formation of {M}axwellian tails.
\newblock {\em Phys. Rev. Lett.}, 36(19):1107--1109, 1976.

\bibitem{bk:GrigorievMeleshkoSuriyawichitseranee[2014]}
Yu. N.Grigoriev, S.~V. Meleshko, and A.Suriyawichitseranee.
\newblock On group classification of the spatially homogeneous and isotropic
  {B}oltzmann equation with sources ii.
\newblock {\em International Journal of Non-Linear Mechanics}, 61:15--18, 2014.

\bibitem{bk:Bobylev[1975a]}
A.~V. Bobylev.
\newblock The method of {F}ourier transform in the theory of the {B}oltzmann
  equation for {M}axwell molecules.
\newblock {\em Dokl. Akad. Nauk SSSR.}, 225:1041--1044, 1975.

\bibitem{bk:Bobylev[1976]}
A.~V. Bobylev.
\newblock On one class of invariant solutions of the {B}oltzmann equation.
\newblock {\em Dokl. AS USSR}, 231(3):571--574, 1976.

\bibitem{bk:GrigMel[1987]}
Yu.~N. Grigoriev and S.~V. Meleshko.
\newblock Group analysis of the integro--differential {B}oltzman equation.
\newblock {\em Dokl. AS USSR}, 297(2):323--327, 1987.

\bibitem{bk:GrigMel[1986]}
Yu.~N. Grigoriev and S.~V. Meleshko.
\newblock Investigation of invariant solutions of the {B}oltzmann kinetic
  equation and its models, 1986.
\newblock Preprint of Institute of Theoretical and Applied Mechanics.

\bibitem{bk:Meleshko[2005]}
S.~V. Meleshko.
\newblock {\em Methods for Constructing Exact Solutions of Partial Differential
  Equations}.
\newblock Mathematical and Analytical Techniques with Applications to
  Engineering. Springer, New York, 2005.

\bibitem{bk:GrigorievIbragimovKovalevMeleshko2010}
Yu.~N. Grigoriev, N.~H. Ibragimov, V.~F. Kovalev, and S.~V. Meleshko.
\newblock {\em Symmetries of integro-differential equations and their
  applications in mechanics and plasma physics}.
\newblock Lecture Notes in Physics, Vol. 806. Springer, Berlin / Heidelberg,
  2010.

\end{thebibliography}
\end{document}